\documentclass[preprint]{aastex} 
\usepackage{amsmath, amsfonts, amssymb, multicol, graphics, pdflscape}
\usepackage[latin1]{inputenc}
\usepackage{tikz}
\usetikzlibrary{shapes,arrows, shadows, positioning}
\usepackage{lscape}
\usepackage[hyphens]{url}
\usepackage{hyperref}

\begin{document}

\title{\ A Mid-infrared Study of RR Lyrae Stars with the WISE All-Sky Data Release}

\author{Tatyana Gavrilchenko\thanks{Astronomy Department, University of California, Berkeley, CA 94720} , Christopher R. Klein\footnotemark[1] , \\Joshua S. Bloom\footnotemark[1] , Joseph W. Richards\thanks{Lawrence Berkeley National Laboratory, Berkeley, CA 94720}}

\begin{abstract} We present a group of 3740 previously identified RR Lyrae variables well-observed with the Wide-field Infrared Survey Explorer (WISE). We explore how the shape of the generic RR Lyrae mid-infrared light curve evolves in period-space, comparing light curves in mid-infrared and optical bands. We find that optical light curves exhibit high amplitudes and a large spectrum of light curve shapes, while mid-infrared light curves have low amplitudes and uniform light curve shapes. From the period-space analysis, we hope to improve the classification methods of RR Lyrae variables and enable reliable discovery of these pulsators in the WISE catalog and future mid-infrared surveys such as the James Webb Space Telescope (JWST). We provide mid-infrared templates for typical RR Lyrae stars and demonstrate how these templates can be applied to improve estimates of mid-infrared RR Lyrae mean magnitude, which is used for distance measurement. This method of template fitting is particularly beneficial for improving observational efficiency. For example, using light curves with observational noise of 0.05 mag, we obtain the same level of accuracy in mean magnitude estimates for light curves randomly sampled at 12 data points with template fitting as with light curves randomly sampled at 20 data points with harmonic modelling. \\

\noindent {\it Subject headings:} infrared: stars; methods: statistical; RR Lyrae: distance scale
\\ \\ Submitted for publication in MNRAS, December 2013.
\end{abstract}

\section{Introduction}\label{intro} 

RR Lyrae stars are pulsating variables that lie near the lower luminosity edge of the instability strip on the Hertzsprung-Russell diagram. At optical wavelengths their mean flux magnitude is nearly constant with a slight metallicity dependance \citep{2004rrls.book.....S, postHipp..book}. At near- and mid-infrared wavelengths, however, RR Lyrae stars exhibit a period-luminosity relation \citep{1986MNRAS.220..279L}. Historically, RR Lyrae stars have been used as important distance indicators for Galactic globular clusters \citep{1939Obs....62..104O,1991ApJ...367..528S, 1976A&AS...25..303C}. More recently, and with the calibration of their longer wavelength period-luminosity relations, RR Lyrae are being used to trace galactic structure near the Galactic Center \citep{2011AcA....61....1S} and in the halo \citep{2010ApJ...708..717S}, as well as map the depth structure of the Magellanic Clouds \citep{2009AcA....59....1S, 2010AcA....60..165S}. An overarching goal of this field is the improvement in RR Lyrae distance measurements through reduced uncertainty in period-luminosity relations. There have been recent advances in the calibration of period--luminosity of Cepheid variable stars \citep{2011ApJ...730..119R, 2012ApJ...758...24F}. 

RR Lyrae stars are evolved low-mass ($\approx 0.7~\rm{M}_\odot$) Population II stars that inhabit the Galactic halo and bulge, and permeate the disk \citep{2004rrls.book.....S}. Their abundance and age make them, in principle, good tracers of Galactic structure. RR Lyrae stars are divided into subclasses based on the shapes of their optical light curves. RRab stars have periods mostly ranging from 0.3 to 0.9 day, {\it V}-band amplitudes of 0.8 to 1.5 magnitudes, and asymmetrical light curves, whereas RRc stars have periods mostly ranging from 0.18 to 0.5 day, {\it V}-band amplitudes of 0.3 to 0.8, and more sinusoidal light curves \citep{1902AnHar..38....1B}. The main difference between the two subclasses is that RRab stars  pulsate in the fundamental mode and RRc stars pulsate in the first radial overtone \citep{2007uvs..book.....P}. Two rarer subclasses are RRd stars, which are double-mode pulsators, and RRe stars, which pulsate in the second radial overtone. The distribution of RR Lyrae subclasses in the LMC as measured by the OGLE-III survey is 71\% RRab, 20\% RRc,  4\% RRd, and 5\% RRe \citep{2009AcA....59....1S}.

Probing the sky in infrared wavebands minimises the effects of interstellar dust and allows detection of faint but warm\footnote{A 725 K blackbody peaks at 4 ${\rm \mu m}$.} sources such as asteroids. The Wide-field Infrared Survey Explorer (WISE) satellite \citep{2010AJ....140.1868W} provides full-sky coverage in four mid-infrared wavebands. The survey mission was to cover 99\% of the sky with at least 8 frames in a 6 month interval then to continue to survey until the cryogenic coolant was exhausted. The survey is up to 60 times more sensitive than its predecessor, IRAS (InfraRed Astronomical Satellite), launched in 1983. The {\it Spitzer} Space Telescope has a flux limit 20 times deeper than typical WISE sensitivity \citep{2006AJ....132.2268M, 2010AJ....140.1868W}. However, this sensitivity requires long exposure times and dedicated pointings, making {\it Spitzer} unsuitable for a complete survey of the sky. The {\it W}1 and {\it W}2 WISE passbands are nearly the same as the first two bands on {\it Spitzer}'s IRAC instrument, which allows WISE to supplement catalogs and follow up on observations made by {\it Spitzer} \citep{2006AJ....132.2268M}.

The upcoming successor to some WISE work will be the James Webb Space Telescope (JWST), since all of the WISE bands fall into the 0.6 to 28 $\mu$m range of JWST. JWST will detect faint, distant RR Lyrae, which will be used as distance markers with near- and mid-IR period-luminosity relations, the latter of which have been derived using WISE data \citep{2011ApJ...738..185K, 2013ApJ...776..135M}.

While variable star science was not among the main objectives of the WISE mission, the survey's scanning pattern enables it to detect variable stars with periods less than 2 days. Due to relatively large amplitudes and strong mid-IR emission, RR Lyrae stars are the most common periodic variables detected by WISE, along with eclipsing Algol variables \citep{2012AJ....143..118H}. The negligible effects of dust in mid-IR bands is another source of motivation for this project \citep{1985ApJ...288..618R}.

With multi-band studies of light curve shape and period-luminosity relations imminent, it will be important to obtain the light curve of a source in several bands. However, well--represented light curves of variable stars can be costly to obtain with respect to telescope time. A series of observations must be made, preferably spanning several period cycles. It is also desirable to sample at random positions in the phase of the object to avoid beating the cycle of pulsation with the cycle of observation. Template fitting is a potential method to recover the same amount of light curve information with fewer observations. Templates have been shown to make accurate mean magnitude and metallicity estimates for {\it V}-band light curves of RRab stars \citep{2007A&A...462.1007K}. Accurate mean magnitude and period measurements are crucial for precision distance studies, especially since period-luminosity relations have been derived for RR Lyrae variables in WISE bands \citep{2011ApJ...738..185K}.

The primary goals of the present research are to acquire a representative sample of RR Lyrae light curves in the mid-IR, to contrast the light curve morphology with light curves available in optical bands, and to formulate a set of mid-IR light curve templates that will aid future studies. To begin, a database of RR Lyrae stars previously identified in visible bands was collated and cross-referenced with the WISE database to form a primary sample set. The final WISE RR Lyrae catalog was created from sources for which periods were derived. These periods were used to create mid-IR light curve models, which appear homogenous in shape, losing distinctive features seen in optical light curves. The models were grouped into typical light curve templates which are an intended tool for the recovery of mean magnitude values for poorly-sampled sources. 

In Section \ref{datasec} we describe in detail the construction of our catalog and our WISE data retrieval methods. We discuss the methodology for determining the period of each source in Section \ref{periods}, the process by which we derive fitted model light curves in Section \ref{robust}, and present statics on the sample in Section \ref{stats}. Finally, we derive a set of typical light curve templates in Section \ref{template_analysis}. Our main results are discussed and summarised in Section \ref{conclusion}.

\section{Data}\label{datasec}

\subsection{Building a Database of RR Lyrae}\label{database}
The first goal was to find a large sample of WISE-observed RR Lyrae stars. A database of previously-identified RR Lyrae stars was created, combining information from GCVS, All Sky Automated Survey (ASAS), Simbad, VizieR, and individual papers. For many of the sources in this database the only available data were the coordinates and RR Lyrae classification. When provided, information about the period, distance, subclass, and magnitude for several different wavebands was also stored. If a single source appeared in multiple surveys or papers, information from all relevant surveys was included, with markers indicating contradicting measurements between surveys. The resulting database contains about 17,000 sources, of which about 5,000 sources have documented {\it V}-band periods. A random sample from the database is shown in Table \ref{db_sample}, with references to the original sources given in Table \ref{refs}. Columns not shown in the table are magnitude values for the following bands: {\it U}, {\it B}, {\it R}, {\it I}, {\it J}, {\it H}, {\it K}, and SDSS {\it ugriz} bands.

\begin{deluxetable}{ccccccc}
\tablecolumns{21}
\tabletypesize{\scriptsize}
\tablewidth{0pt}
\tablecaption{Sample of the RR Lyrae Catalog}
\tablehead{\colhead{ID} & \colhead{RA} & \colhead{DEC} & \colhead{Subclass} & \colhead{Period}  & \colhead{{\it V} Magnitude}  & \colhead{Reference}}

\startdata
V* V532 Her & 272.98417 & 28.06606 &  &  &  & 11,17 \\
154584 & 110.1625 & 6.6816667 & RRab & 0.616 &  & 20,22 \\
NSV 06930 & 227.0512 & -69.8428 &  &  &  & 17 \\
V2637 Sgr & 273.1692 & -36.725 & RRab &  & 16.2 & 17 \\
OGLE BUL-SC45 245339 & 18.0595 & -30.029 &  &  & 16.96 & 1 \\
MACHO 176.18702.1147 & 269.9505 & -27.28969 &  &  & 18.6 & 19 \\
V* V4267 Sgr & 272.1705 & -31.54319 & RRab &  & 16.82/17.2 & 3,17,19 \\
V0344 Cam & 66.0633 & 71.6539 & RRc &  & 14.1 & 17 \\
ZZ And & 12.395 & 27.0222 & RRab & 0.555 & 13.34/13.025 & 17,19,20,21,22 \\
278058 & 268.0971 & -29.7703 &  & 0.466 & 19.96 & 15 \\
OGLE LMC-SC6 40925 & 80.07317 & -69.67808 &  &  & 16.79 & 19 \\
MACHO 104.20383.612 & 270.9915 & -27.918 &  & 0.598 & 17.65/17.61 & 15,19 \\
MACHO 134.33910.838 & 278.8755 & -27.43281 &  &  & 16.95 & 19 \\
V* V505 Lyr & 279.14967 & 28.06694 & RRab &  & 15.8/16.425 & 11,17 \\
NSV 11395 & 282.9525 & -56.7917 &  &  &  & 17 \\
MACHO 142.27902.808 & 275.376 & -29.29411 &  &  & 18.57 & 19 \\
276501 & 268.35315 & -30.1274 &  & 0.444 & 18.23 & 15 \\
V3003 Sgr & 274.7667 & -31.4717 & RRab &  & 16.45 & 17 \\
V0759 Cep & 7.3362 & 84.7625 & RRc &  & 15.83 & 17 \\
BB97c 37 & 272.68462 & -31.97056 & RRab & 0.63 & 16.43/16.85 & 4,17,19 \\
V* V1357 Sgr & 270.6475 & -29.9583 &  &  &  & 2 \\
V2466 Sgr & 269.9404 & -28.8722 & RRab & 0.623 & 17.63/17.35/18.55 & 15,17,19 \\
253827 & 280.17039 & -52.762774 & RRab & 0.401 &  & 20,21 \\
V* V4607 Sgr & 272.66654 & -32.10228 & RRab & 0.585 & 17.7 & 4,17 \\
MACHO 139.32982.4416 & 278.3235 & -28.59539 &  &  & 16.65 & 19 \\
2MASS J00112104+0010138 & 2.83767 & 0.1705 &  & 0.563 & 16.54 & 8,19 \\
HL Com & 184.5954 & 30.1883 & RRc &  & 15.7 & 17 \\
MACHO 101.20911.863 & 271.254 & -27.34639 &  &  & 17.36 & 19 \\
T Hyi & 6.0663 & -76.2819 &  &  &  & 17 \\
\enddata
\label{db_sample}
\tablecomments{ Columns showing magnitude for the {\it U}, {\it B}, {\it R}, {\it I}, {\it J}, {\it H}, {\it K}, and SDSS {\it ugriz} bands are omitted.}
\end{deluxetable}

\begin{deluxetable}{cccc}
\tablecolumns{2}
\tabletypesize{\scriptsize}
\tablewidth{0pt}
\tablecaption{References for the RR Lyrae Catalog}
\tablehead{\colhead{Reference Number} & \colhead{Reference} & \colhead{Reference Number} & \colhead{Reference}}
\startdata
1 & \cite{2008AA...481..441G} & 14 & \cite{2005AA...442..381M} \\
2 & \cite{1984AJ.....89.1836B} & 15 & \cite{2006ApJ...651..197C} \\
3 & \cite{1992AJ....103.1872B} & 16 & \cite{2010AJ....139..357Z} \\
4 & \cite{1997AJ....114.2596B} & 17 & \url{http://www.sai.msu.su/gcvs/gcvs/index.htm} \\
5 & \cite{1995MNRAS.275..929C} & 18 & \cite{2009AcA....59...33P} \\
6 & \cite{1982AJ.....87..314K} & 19 & \url{http://simbad.u-strasbg.fr/simbad/} \\
7 & \cite{1965ApJS...11..223K} & 20 & \cite{2012ApJS..203...32R} \\
8 & \cite{2008ApJ...678..865M} & 21 & \cite{1997AcA....47..467P} \\
9 & \cite{1984ApJ...283..580S} & 23 & \cite{2007AA...475.1159D} \\
10 & \cite{1991ApJ...367..528S} & 24 & \cite{2003AcA....53...27P} \\
11 & \cite{1996AJ....112..742W} & 25 & \cite{2002AA...389..149D} \\
12 & \cite{1994AA...291..121S} & 26 & \cite{2007AA...475.1159D} \\
13 & \cite{2007AA...463..589S} \\
\enddata
\label{refs}
\end{deluxetable}

The database inherits the biases of the surveys used to create it, so it is neither complete nor unbiased. In addition, there may be contamination by sources incorrectly classified as RR Lyrae. The most likely contaminants are eclipsing binary systems in $\sim$0.2 day period orbits where the two bodies have similar masses, since these result in sinusoidal light curves similar to those of the RRc subclass. An alternative to this procedure is to select all stars with a variability flag from the WISE database, $\sim$460,000 objects, and to make further colour, period, and amplitude restrictions to form a sample of likely RR Lyrae candidates. However, since the bounds of such cuts are not well-studied in {\it W}1 and {\it W}2--bands, this procedure may result in a sample highly contaminated by Algol-type eclipsing binaries and W UMa binaries.

\subsection{Retrieving Data from WISE}\label{data}
WISE performed an all-sky survey in four mid-infrared bands centred at 3.4, 4.6, 12, and 22 \micron. The strategy of WISE was to scan in great circles with a centre located at the Sun. The result of this scanning pattern was a high number of exposures taken around the ecliptic poles and a low number of exposures taken near the ecliptic plane. Typically, points around the ecliptic poles were observed over 100 times and points around the ecliptic plane were observed 12 to 13 times. The full cryogenic mission phase began on 7 January 2010, completed full coverage of the sky on 17 July 2010, and surveyed approximately 20\% of the sky a second time before the mission phase ended on 6 August 2010. A preliminary data release with coverage of 57\% percent of the sky was released in April 2011, superseded by the all-sky data release in March 2012\footnote{\url{http://wise2.ipac.caltech.edu/docs/release/allsky/}}.

\begin{figure}[h!]
\begin{center}
\includegraphics[width=1\textwidth]{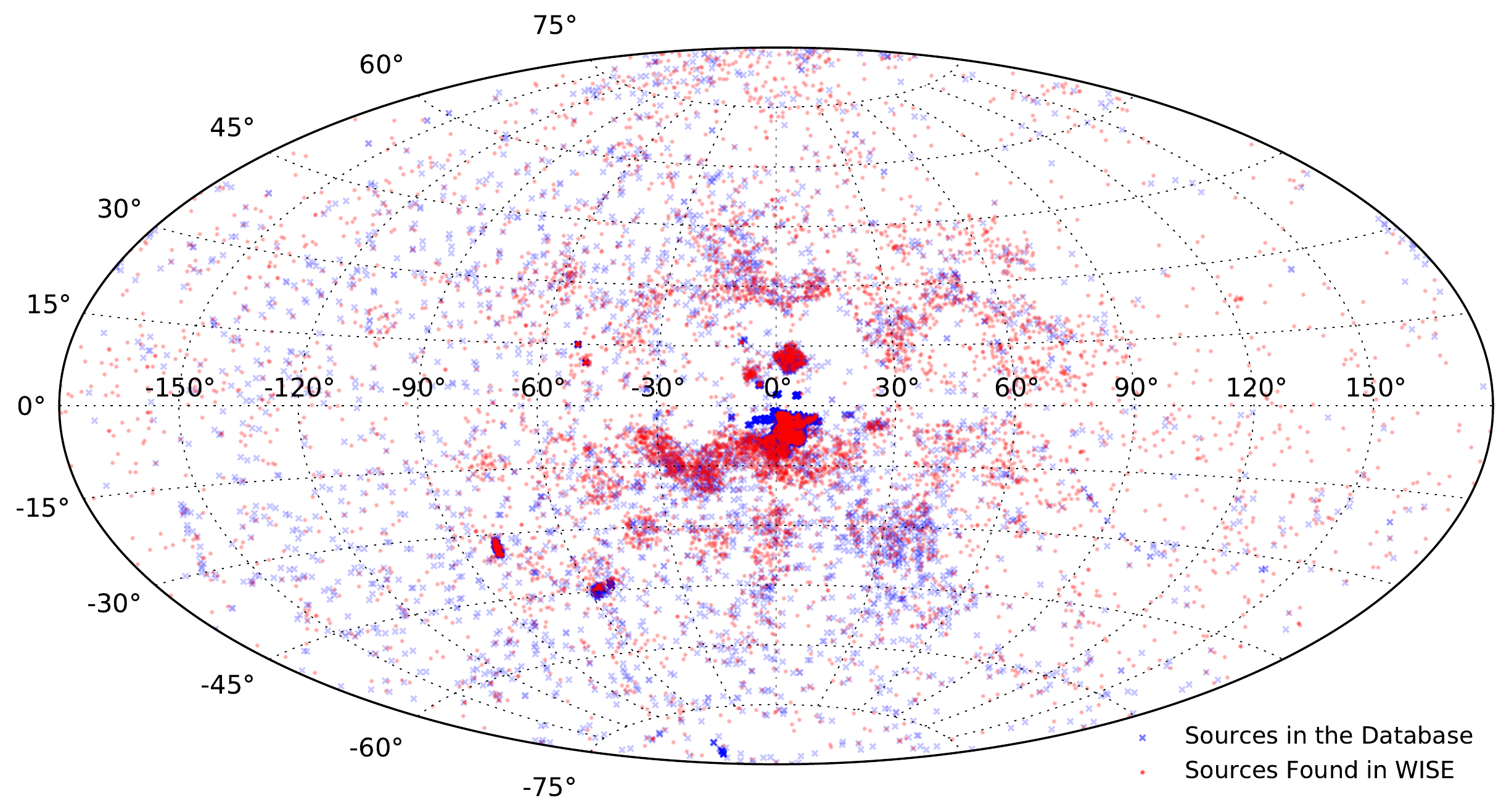}
\caption{The distribution of the $\sim$8,500 RR Lyrae sources found in the WISE catalog (red circles) superimposed on the $\sim$17,000 sources of the RR Lyrae database (blue crosses). The Aitoff plot coordinate system is in Galactic coordinates.  \label{spatialfig}}
\end{center}
\end{figure}

Coordinates from the compiled database of known RR Lyrae stars were cross-referenced with the WISE all-sky release catalog. A search radius of 10 arcseconds was used and data points with artefact contamination flags or photometric quality flags were rejected. The coordinate-linked cross-reference search returned $\sim$8,500 sources with ten or more measurements in either the {\it W}1 or {\it W}2 bands. Analysis of the {\it W}3 and {\it W}4 light curves is omitted, since both the photometric quality and the number of epochs in those longer-wavelength bands are greatly diminished compared to {\it W}1 and {\it W}2. Figure \ref{spatialfig} is a spatial plot of the RR Lyrae database, showing the sources with matched WISE light curves in red. 

Despite choosing the highest quality photometric flags, some of the light curves contained outliers. The variable nature of these stars makes outlier rejection tricky. To exclude outlying points, any observations located 0.5 or more magnitudes away from the mean light curve magnitude were eliminated. This threshold was chosen because it is well above the mid--infrared peak--to--peak amplitude for RRab stars, which was computed to be 0.225 magnitudes in Section \ref{stats}. Outlier rejection was done recursively by removing the point farthest from the mean magnitude if it was over 0.5 magnitudes away, then recalculating the mean magnitude and repeating the procedure.

\section{Mid-IR Light Curves}

\subsection{Determining Periods}\label{periods}
WISE light curves in the {\it W}1 and {\it W}2 bands were used to attempt to determine a period for each source. The two period-finding algorithms used were  Lomb-Scargle \citep{lomb76, barning63, scargle82} and Super-Smoother \citep{Reimann}. The Lomb-Scargle algorithm uses a least-squares frequency analysis of a range of test frequencies to create a periodogram, which is an estimate of the power spectrum. A modified Lomb-Scargle algorithm developed by \cite{2011ApJ...733...10R} was used in the present work. While the original algorithm fixes the mean magnitude of the model light curve at zero, the modified algorithm allows a floating mean value. The reason for this alteration is that observations of a source which are not uniformly sampled in period space will result in an inaccurate mean magnitude measurement. Since equally-spaced measurements in phase space are not expected with WISE data, this modification results in more robust period estimates. The Lomb-Scargle algorithm outputs the period, the periodogram, and the harmonic model, using up to 8 harmonic components. The {\it Supersmoother} algorithm folds the light curve on a range of test periods and determines which test periods result in the smoothest folded light curve. The outputs are the period for the source, a goodness of fit of the model to the data, and the epoch of maximum brightness. Both algorithms require a light curve as the input, which is a list of the timestamp, magnitude, and the magnitude error for each observation of the source. 

The preliminary list of candidate periods was comprised of the top three periods returned from each of the two period-finding algorithms, so six total candidate periods for each of the two bands. These lists of candidate periods were trimmed by removing periods outside the 0.2 to 1 day range of RR Lyrae variables and by averaging any two candidate periods for a given source which were less than 0.05 day apart. This threshold for unique periods was used because the scanning pattern WISE completes one orbit in 90 minutes, which corresponds to 0.0625 day. The final lists typically had between one and three candidate periods for each source per waveband.

If possible, a single period from the candidate list was chosen as the most likely true period, which means that the periodogram attains a maximum at that period. For a light curve with few magnitude measurements and high photometric errors, the periodogram may have several peaks, making it difficult to identify the true period. For these ambiguous cases, no true period can be determined and these sources are excluded from further analysis. A period-determining process that minimises such occurrences is desired. The schematic in Figure \ref{schematic} is a representation of the process used, and the text in this section describes each step in detail.

\tikzstyle{tiny} = [rectangle, text width=1em, text centered, rounded corners, minimum height=1em]
\tikzstyle{largest_block} = [rectangle, draw, drop shadow,fill=white, text width=15em, text centered, rounded corners, minimum height=4em]
\tikzstyle{smaller_block} = [rectangle, draw, drop shadow,fill=white, text width=5em, text centered, rounded corners, minimum height=4em]
\tikzstyle{blue_oval} = [rectangle, draw, drop shadow,fill=blue!10, text width=7em, text centered, rounded corners, minimum height=4em]
\tikzstyle{red_oval} = [rectangle, draw, drop shadow,fill=red!10, text width=7em, text centered, rounded corners, minimum height=4em]
\tikzstyle{line} = [draw, -latex', align=center]
\tikzstyle{med_block} = [rectangle, draw, drop shadow,fill=white, text width=9em, text centered, rounded corners, minimum height=4em]
\tikzstyle{large_block} = [rectangle, draw, drop shadow,fill=white, text width=10em, text centered, rounded corners, minimum height=4em]
\tikzstyle{long} = [rectangle, draw,  drop shadow,fill=white, text width=13em, text centered, rounded corners, minimum height=2em]

\begin{figure}[h!!]
\begin{tikzpicture}[thick,scale=0.8, every node/.style={transform shape}]

\node [tiny] (initial) { };
\node [label={[label distance=0.15cm]0:7945 sources}][long, left of = initial] [node distance = 5.5cm] (w1_cand) {initial {\it W}1 candidate list};
\node [label={[label distance=0.15cm]0:5248 sources}][long, right of = initial] [node distance = 5.5cm] (w2_cand) {initial {\it W}2 candidate list};
\node [label={[label distance=0.15cm]0:3688 sources}][long, below of = w1_cand] [node distance = 1.5cm] (w1_cuts) {quality cuts};
\node [label={[label distance=0.15cm]0:1857 sources}] [long, below of = w2_cand] [node distance = 1.5cm] (w2_cuts) {quality cuts};
\node [long, below of = w1_cuts] [node distance = 1.5cm] (w1_dups) {duplicate removal};
\node [long, below of = w2_cuts] [node distance = 1.5cm] (w2_dups) {duplicate removal};
\node[largest_block, below of = initial, node distance=4.5cm] (check) {How many periods appear in both lists (within 0.05 day)};
\node [label={[label distance=0.15cm]0:1333 sources}][smaller_block, below of=check, node distance=2.5cm] (one) {one};
\node [label={[label distance=0.15cm]0:6923 sources}][smaller_block, left of=one, node distance=5.3cm] (zero){none};
\node [label={[label distance=0.15cm]0:121 sources}][smaller_block, right of=one, node distance=5.3cm] (two){two or more};
\node [label={[label distance=0.15cm] 270:1333 sources}][blue_oval, below of=one, node distance=2.5cm] (done1) {Accept Average of the {\it W}1 and {\it W}2 Periods};
\node [med_block, below of=two, node distance=2.5cm] (highestPSD){Find the pair of periods with highest {\it W}1 PSD peak};
\node [label={[label distance=0.15cm]270:104 sources}][blue_oval, below of=highestPSD, node distance=2.5cm] (done2) {Accept Average of the {\it W}1 and {\it W}2 Periods};
\node [largest_block, below of = zero, node distance=2.5cm] (high_psd) {Take the final candidate for each band to be the period with the highest PSD in that band};
\node [med_block, below of = high_psd, node distance=2.5cm] (num_bands) {For how many bands does a final candidate exist?};
\node [tiny, below of = num_bands, node distance=2.5cm] (holder) { };
\node [red_oval, left of= holder, node distance=3cm] (check_dist) {Reject Source};
\node [label={[label distance=0.15cm]270:2094 {\it W}1 sources}] [label={[label distance=0.7cm]270: 263 {\it W}2 sources}] [blue_oval, right of=holder, node distance=3cm] (accept) {Accept Period for that Band};

\path [line] (w1_cand) -- (w1_cuts);
\path [line] (w2_cand) -- (w2_cuts);
\path [line] (w1_cuts) -- (w1_dups);
\path [line] (w2_cuts) -- (w2_dups);
\path [line] (w1_dups.south) |- (check.177);
\path [line] (w2_dups.south) |- (check.3);
\path[line] (check) -- (one);
\path[line] (check.357) -| (two);
\path[line] (check.183) -| (zero);
\path [line] (one) -- (done1);
\path [line] (two) -- (highestPSD);
\path [line] (highestPSD) -- (done2);
\path [line] (zero) -- (high_psd);
\path [line] (high_psd) -- (num_bands);
\path [line] (num_bands.0) -| node[right = .15 cm, pos = 0.8] {One Band} (accept);
\path [line] (num_bands.180) -| node[left = .15 cm, pos = 0.8] {Both Bands} (check_dist);

\end{tikzpicture}
\caption{A schematic of the process used for period determination from a list of candidate periods. See section \ref{periods} for more detailed explanation of the process.}
\label{schematic}
\end{figure}
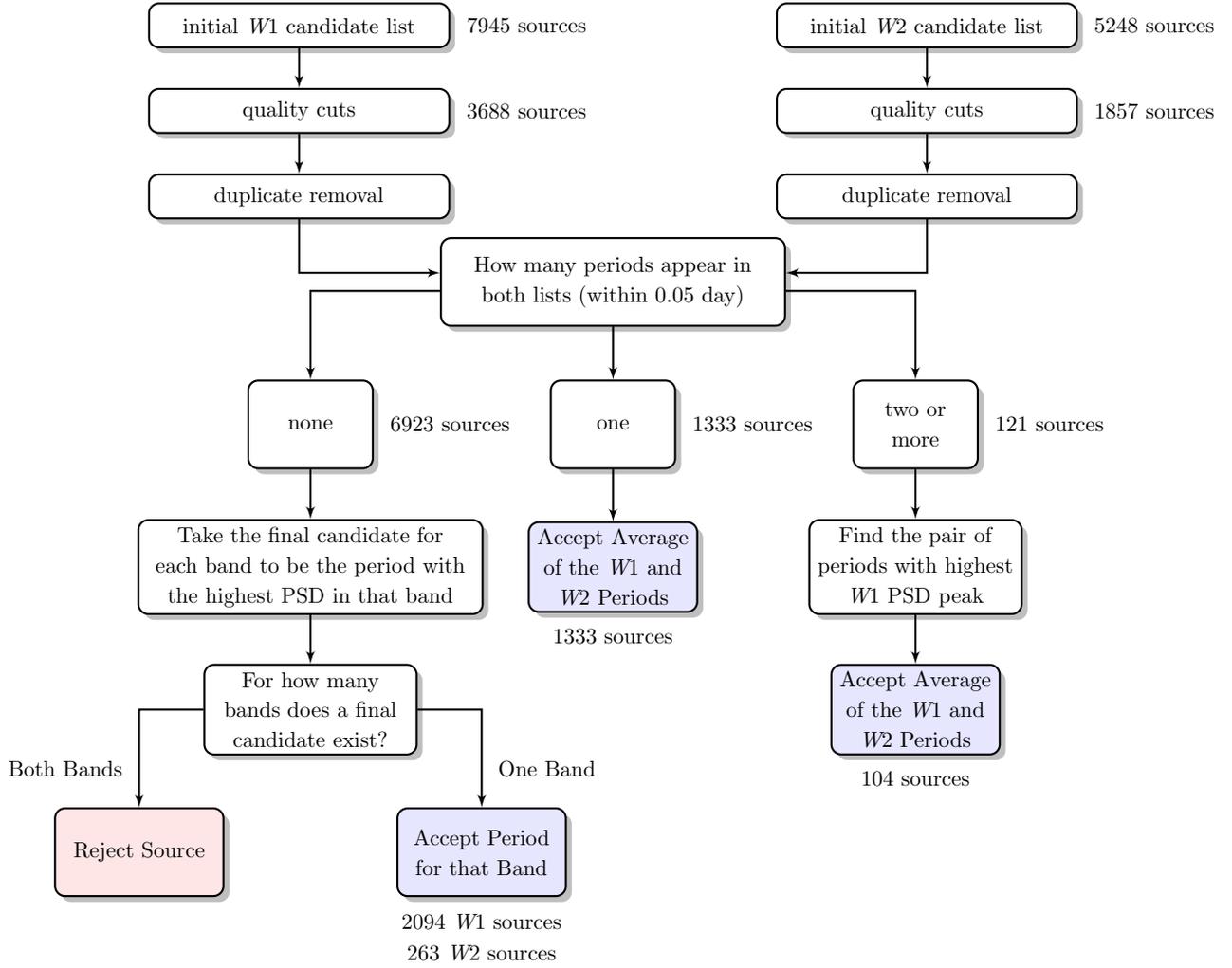

A harmonic model for each of the candidate periods was created, once again using the Lomb-Scargle algorithm, but now restricting the interval of scanned periods to be centered at the candidate period and to have a width of 0.1 day (effectively boosting the periodogram resolution). In addition, points in the original light curve which were over four standard deviations away from the model light curve were considered outliers and eliminated from further investigation. Upon running the Lomb-Scargle process, several features of the harmonic model were measured to serve as quality markers of the model. These features are the period, the periodogram peak height, the residual determined by the mean of the squares of the errors between the data and the model, the number of observations in a light curve, the largest gap in phase space of the folded light curve, the ratio of the data amplitude and the model amplitude, the average photometric error divided by the amplitude of the model, and the number of critical points in the model. The initial quality cuts of the period candidate selection are:

\begin{itemize}
\item candidate period is between 0.2 and 1 day
\item periodogram peak power spectral density  is greater than 3 dB Hz$^{-1}$
\item harmonic residual is less than 0.006 magnitudes
\item maximum gap in phase space is less than 30\% of a period cycle
\item the ratio of data amplitude to model amplitude is greater than 0.8 and less than 2.5
\item the ratio of average photometric error to model amplitude is less than 0.3
\item the model has only two critical points (where the derivative changes sign)
\end{itemize}

The final condition requiring exactly two critical points is reasonable because an RR Lyrae light curve is expected to have only one local maximum and one local minimum in a single pulsation cycle. If more than two critical points are detected, it is an indication of a poor model fit or an incorrect period estimate. For example, if a light curve is folded on a period which is twice the value of the true period, then two pulsation cycles will be shown. To eliminate multiples of the true period from consideration, a candidate light curve does not pass through the quality cuts if it has more than two critical points. To allow room for regions of relative flatness in the light curve model, any neighbouring critical points whose difference is less than 20\% of the total model amplitude are considered to be a single critical point. This condition is strong enough to reject models that are clearly not RR Lyrae stars, yet allows leeway for borderline cases. 

Next, a second duplicate removal was performed on each candidate list: if two periods less than 0.05 days apart were found in a list, the period with the higher PSD peak was kept and the other period was removed. The next step was to check if a single period is present in both the {\it W}1 and {\it W}2 shortened candidate lists. This is motivated by the expectation of an RR Lyrae star to have a single period regardless of the bandpass used to observe it. Periods appearing in both candidate lists up to an uncertainty of 0.05 day were flagged as highly likely to be the true period. If there was exactly one pair of flagged periods, the average of the {\it W}1 and {\it W}2 periods was taken to be the true period for both bands. If there were two or more pairs of flagged periods, then the pair of periods with the highest {\it W}1 PSD peak was averaged and taken to be the true period. If no periods were flagged, then the lists are once again considered separately.

Next the period with the highest PSD peak for each band was marked as the final candidate. If both bands contained a final candidate, the two candidates must be farther than 0.05 day apart, since otherwise they would have been flagged in the previous step. Since it is forbidden for a single source to have two distinct periods, the source must be rejected from the database. If a final candidate exists for only one band, then that period is taken to be the true period for the source. 

It is important to note that both period-finding algorithms returned a surprisingly high number of RR Lyrae stars with periods between 0.1 and 0.2 day. Oftentimes, a candidate list had a low period in this range along with a more typical period in the 0.4 to 0.6 day range. These two candidate periods resulted in harmonic models with very similar harmonic residuals. Since WISE completes one orbit in 90 minutes, which is 0.0625 day, the time between successive observations is most often that value. This results in an observation bias that may confuse the period-folding algorithms into favouring periods shorter than 0.2 days. To avoid this bias, the period range of RR Lyrae in the catalog was restricted to be between 0.2 and 1 day. 

\subsection{Creating Robust Models}
\label{robust} 
\begin{figure}[h!]
\begin{center}
\includegraphics[width=0.82\textwidth]{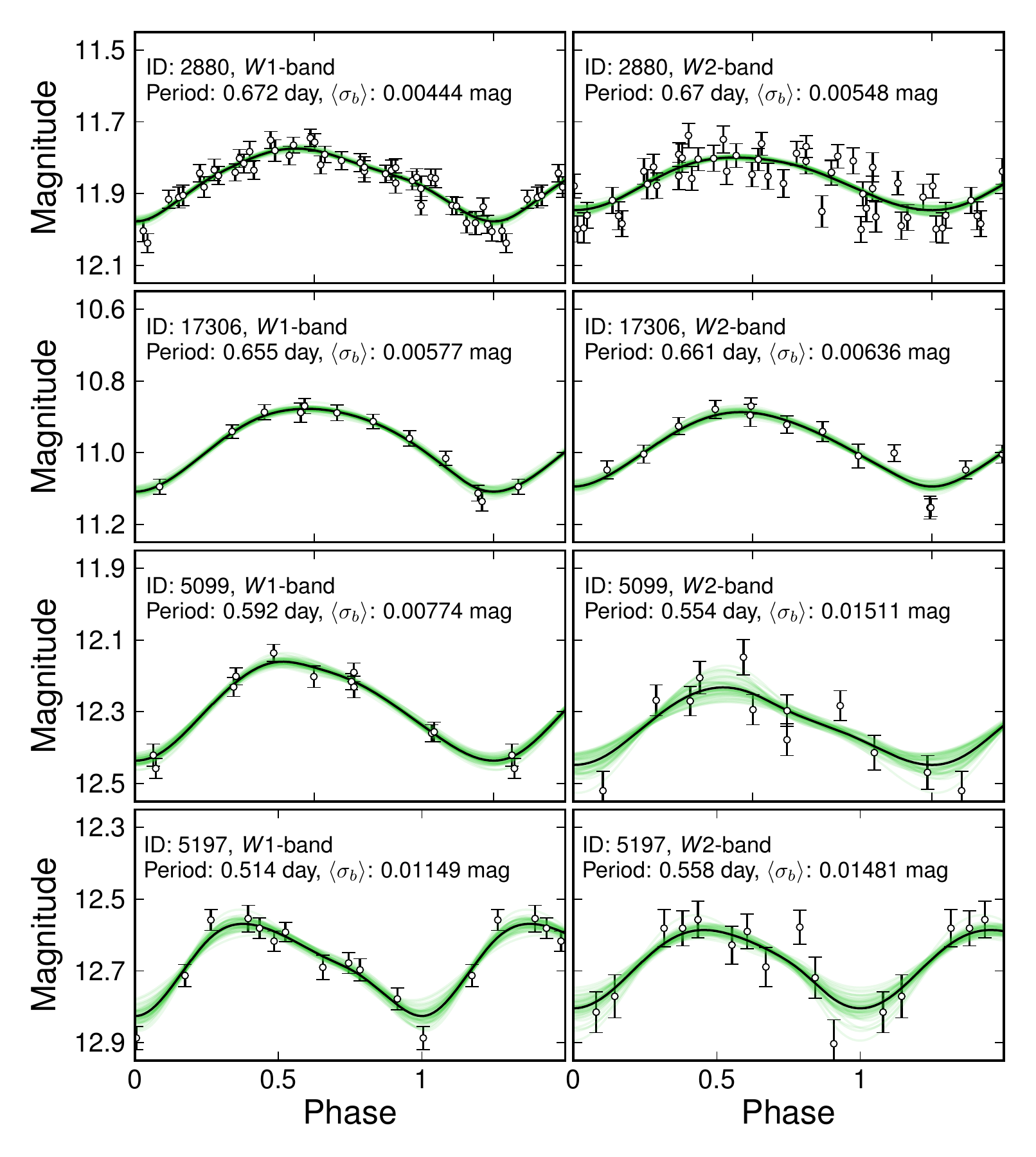}
\caption{
Several sources from the catalog of WISE RR Lyrae stars are shown, with the {\it W}1-band on the left and the {\it W}2-band on the right. For the bootstrapping process, 500 perturbed models are generated by perturbing the magnitude of each data point and fitting a harmonic model with 8 harmonics to each new set of points. 50 of these models are drawn as green curves. The robust model, drawn as a black curve, is the average of the perturbed models. This model is used in all further analysis. The standard deviation of the perturbed models, $\sigma_{b}$,  gives an error for the robust model as a function of phase. The average standard deviation over one cycle of the light curve, $\langle \sigma_{b} \rangle$, is an effective quality measure of the light curve. \label{bootstrap}}
\end{center}
\end{figure}

To remove fluctuations in the light curve shape that are actually instrumental inconsistencies and random errors in photometry, a bootstrap method was used on each of the sources for which a period was determined. In addition, this process better incorporates error into the model. For a single source, the magnitude of each data point in the raw light curve was perturbed. The perturbed magnitude was chosen randomly from a Gaussian distribution of possible magnitudes centred at the original magnitude with a standard deviation equal to the photometric error of that point. Once each data point was perturbed in this somewhat random way, a perturbed harmonic model was produced once again using the Lomb-Scargle method, fixing the period to a range of 0.01 day centred at the derived period. This process was repeated 500 times, and the average of these 500 perturbed models is referred to as the robust model. In addition, the standard deviation of the perturbed models was calculated to obtain an error measurement for the robust model, which is referred to as the robust model error. Four examples of robust models in both bands are shown in Figure \ref{bootstrap}. This process provides an error measurement at any point in the model light curve, and the robust model error serves as an assessment of the overall quality of the model. 

If the ratio of the robust model error to the model amplitude for the {\it W}1- or {\it W}2-band light curve is no greater than 0.02, the source is added to the final catalog of WISE RR Lyrae stars. The resulting catalog contains 3740 sources with {\it W}1 light curve models and 1584 sources with {\it W}2 light curve models. Of these, 1424 sources have light curve models in both bands. The average number of epochs is 18 for a source in {\it W}1 and 19 for a source in {\it W}2. The average mean magnitude and average amplitude and 12.468 mag and 0.220 mag for {\it W}1 and 11.417 mag and 0.216 mag for {\it W}2. Distances and distance uncertainties were calculated using mid--infrared period--luminosity relations \citep{chris.in.prep}. These period--luminosity relations are dependent on the RR Lyra subclass. If a source had been previously classified, the previous classification was used. Otherwise, sources with a period less than 0.4 day were tentatively classified as RRc and sources with periods greater than 0.4 day were classified as RRab. If a source had a distance value in both the {\it W}1- and {\it W}2-band, these two distances were averaged and the uncertainties propagated accordingly. The final catalog of WISE RR Lyrae stars is available in electronic format as supplemental material.

\section{Characterisation of RR Lyrae in the Mid-IR}\label{stats}

To provide a comparison for mid-IR light curves, a sample of optical light curves of RR Lyrae stars was obtained from the OGLE-III Catalog of Variable Stars \citep{2009AcA....59....1S}. The sample contains 23,920 RRab, RRc, and RRe stars in the LMC, with a distribution of 74\% RRab, 21\% RRc, and 5\% RRe. Here, RRe stars are defined as low amplitude RR Lyrae stars with periods of less that 0.3 day, and the distribution between RRc and RRe stars is more or less continuous (see Figure 3 in \citeauthor{2009AcA....59....1S}). Since the distinction is difficult to make with WISE data, for all further discussion stars classified as RRe by the OGLE-III Catalog are grouped in the RRc subclass. Using this convention, the OGLE-III LMC sample has a distribution of 74\% RRab and 26\% RRc. The LMC yields a large, complete, and uniform sample and the effects of Galactic dust are minimised. The two filters used are {\it V}-band centred at 551 nm and {\it I}-band centred at 806 nm, the latter of which may be considered to probe near-infrared wavelengths.

Harmonic light curves for the OGLE-III sample were created using a methodology similar to the creation of the WISE harmonic models. First, the light curve for each source was perturbed 100 times as in Section \ref{robust}. The Lomb-Scargle method was used to find the period with the highest power spectral density within a range of 0.01 day centred at the period provided by OGLE. A harmonic model with up to 8 harmonics was then computed for this period. The average of the 100 perturbed harmonic models was taken to be the robust model. To ensure that these models fit the light curve well, a model rejection process was run based on the percentage of outlying light curve points. Since the {\it V}-band light curves tended to have far fewer measurements and far smaller measurement errors than {\it I}-band light curves, different outlier conditions were used for the two bands. {\it V}-band light curves with over 10\% outlying data points were rejected, where an outlying point was defined to be over 8 times its measurement error from the model. {\it I}-band light curves with over 5\% outlying data points were rejected, where an outlying point was defined to be over 4 times its measurement error from the model. Note that this procedure eliminated most of the RR Lyrae undergoing Blazhko--modulation from the sample, since these light curves are characterised by a spread in the light curve shape, likely resulting in more outliers. 

\begin{figure}[h!]
\begin{center}
\includegraphics[width=0.75\textwidth]{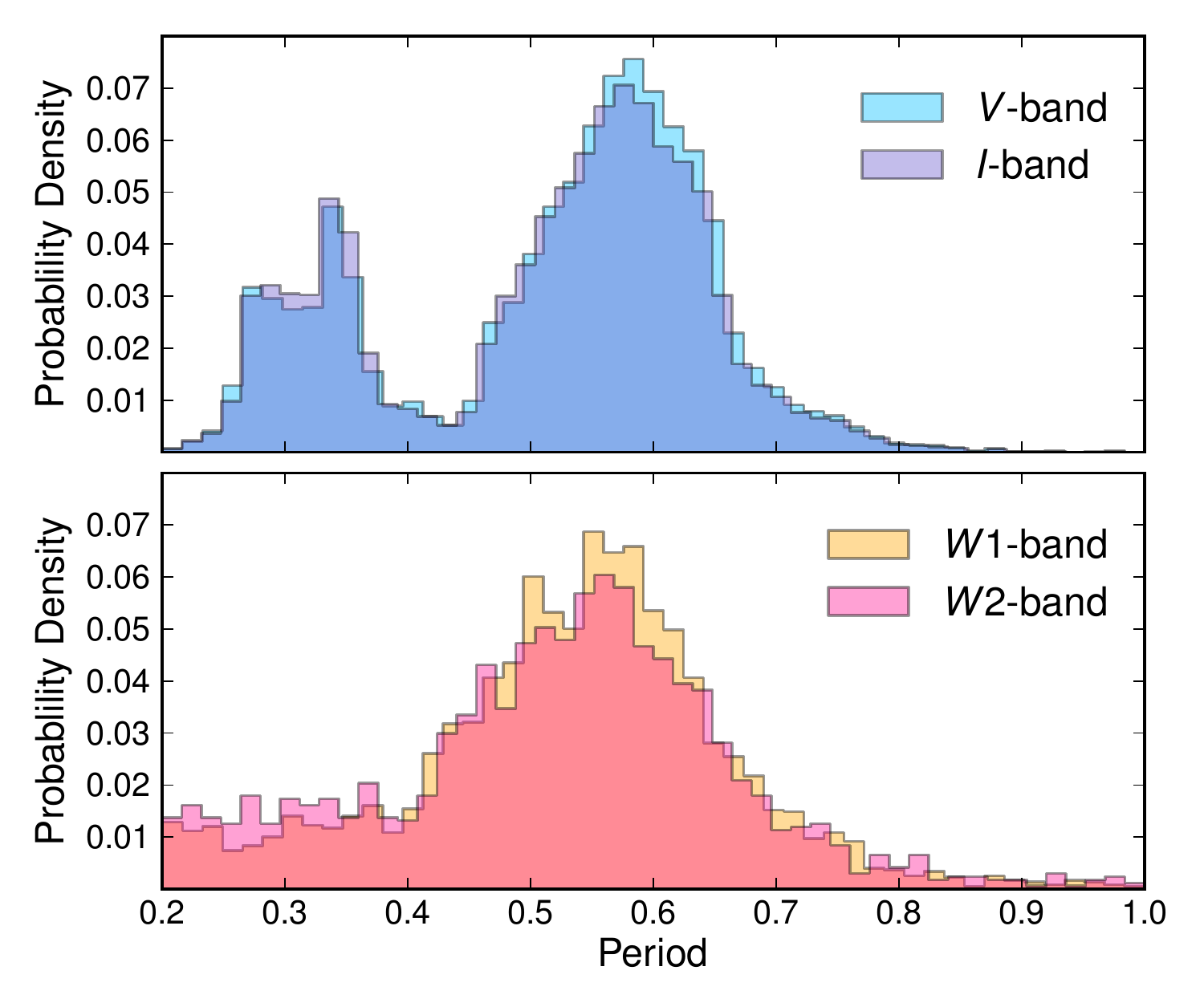}
\caption{Histograms of periods for the full WISE and OGLE databases. For the WISE period--finding process, candidate periods were restricted to be between 0.2 and 1 day. The y-axis plots the probability density of period bin, which will be 1 when integrated over all period bins. \label{hist}}
\end{center}
\end{figure}

The final optical-comparison sample is comprised of $\sim$13,000 {\it V}-band light curve models, with 78\% RRab and 22\% RRc, and $\sim$18,500 {\it I}-band light curve models, with 75\% RRab and 25\% RRc. A histogram of the periods in the full WISE and OGLE catalogs is shown in Figure \ref{hist}. In the figure, the secondary peak representing subclass RRc stars that occurs between 0.25 and 0.4 day in optical wavelengths is missing from the mid-infrared histogram. This occurs because subclass RRc stars have lower amplitudes than RRab stars but similar photometric errors, and the quality cuts in the light curve selection process restricted the ratio of average photometric error to model amplitude to be greater than 0.2. Thus, many RRc sources were omitted from the sample due to observational limitations. 

\begin{figure}[h!]
\begin{center}
\includegraphics{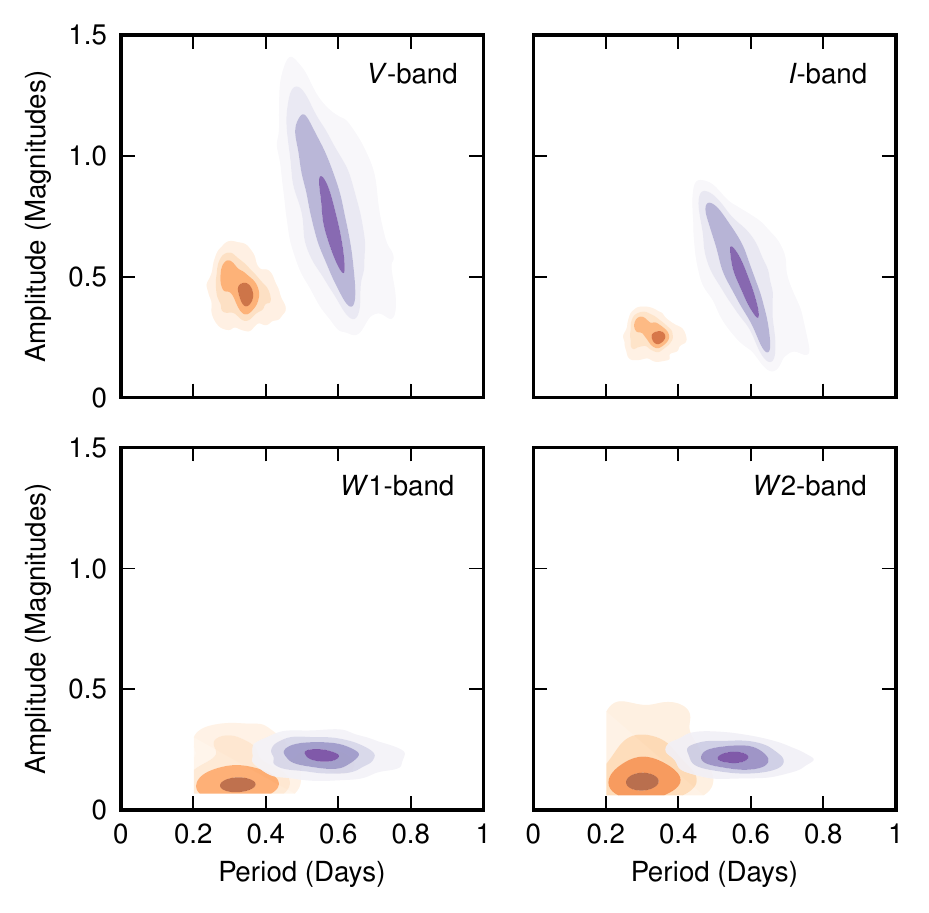}
\caption{Contour plot period--amplitude relations for RR Lyrae stars in four wavebands. RRab subclass stars are denoted by the purple contours and RRc subclass stars are denoted by the orange contours. Contour lines are drawn independently for each subclass and have levels at the 99, 95, 90, 80, and 70th percentile. The data shown in the two upper panels are a subset of the OGLE-III Catalog of Variable Stars in the LMC. The {\it V}-band sample contains 10,101 RRab and 2,690 RRc and the {\it I}-band sample contains 13,761 RRab and 4,203 RRc sources (see section \ref{stats} for details on the sample selection process). The two lower panels are a subset of the final WISE RR Lyrae catalog, showing only sources which have subclass identification from previous surveys, which is 38\% of the catalog. The {\it W}1-band subset contains 2,362 RRab and 288 RRc sources  and the {\it W}2-band subset contains 1,229 RRab and 187 RRc sources. \label{per_amp}} 
\end{center}
\end{figure}

One prominent difference between RR Lyrae stars in optical bands and mid-IR bands is that the mean amplitude decreases as the wavelength increases. Figure \ref{per_amp} depicts period--amplitude relations in the four wavebands: the OGLE {\it V}- and {\it I}-bands, and the {\it W}1 and {\it W}2-bands. Note that the 0.2 day cutoff in period in the two WISE plots is directly induced (see the end of Section \ref{periods}), while the 0.1 amplitude cutoff can be explained as an artefact of the fixed mean error-to-amplitude ratio. This suggests that there exist RR Lyrae stars with even smaller {\it W}1- and {\it W}2-band amplitudes, but these are difficult to detect with WISE since the average photometric error is on the order of the amplitude of the light curve. In addition, the WISE plots only show the sources in the final WISE RR Lyrae database that have been classified into the RRab or RRc subclasses by previous surveys, which is 38\% of the catalog. While all four plots present a distinction between the RRab and RRc subclasses, these subclasses are more distinguished in the {\it V}- and {\it I}-bands and they begin to blend together in the {\it W}1- and {\it W}2-bands. The average amplitude is 1.147 for RRab and 0.546 for RRc in the {\it V}-band, 0.595 for RRab and 0.279 for RRc in the {\it I}-band, 0.225 for RRab and 0.156 for RRc in the {\it W}1-band, and 0.22 for RRab and 0.19 for RRc in the {\it W}2-band. These tight period--amplitude relations in mid-IR bands may be useful for identification and classification purposes. 

\begin{figure}[h!]
\begin{center}
\includegraphics[width=1 \textwidth]{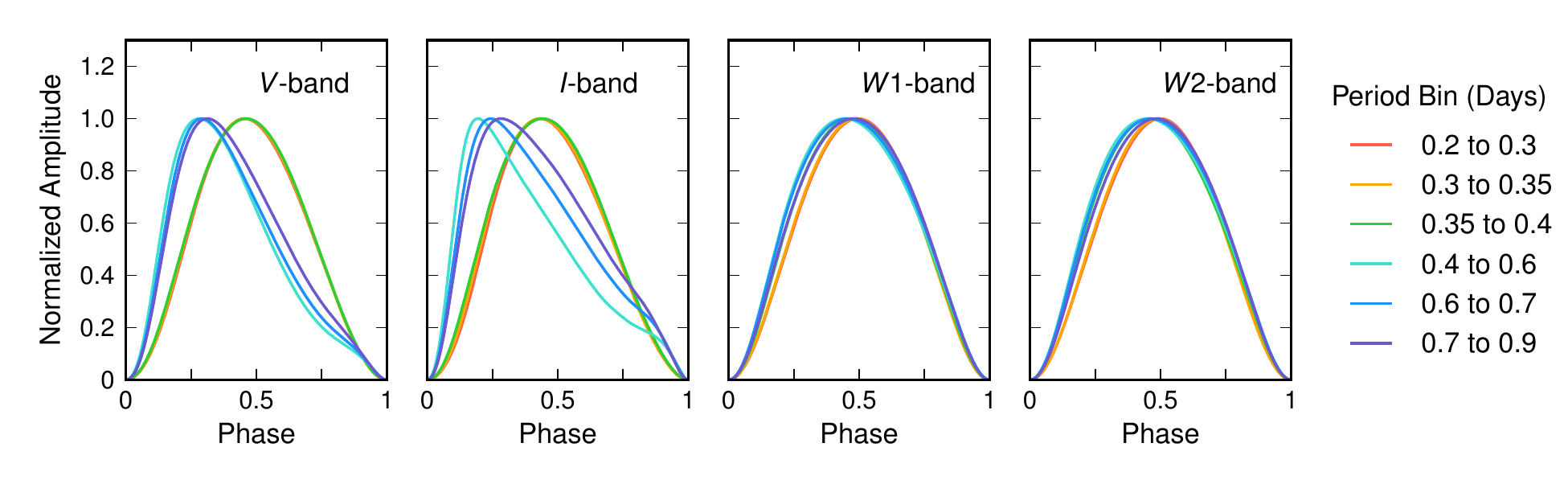}
\caption{Comparison of light curve morphology in optical and infrared bands. Optical light curves are obtained from {\it I}-band and {\it V}-band photometry of subclass RRab and RRc RR Lyrae in the LMC from the OGLE-III Catalog of Variable Stars. Light curve models have been normalised to be 0 at maximum magnitude and 1 at minimum magnitude. Light curves have been grouped into 6 bins according to their period. All light curves in a period bin were averaged to create the coloured curves. In the {\it V}- and {\it I}-bands there is notable variation between light curves with periods less than 0.4 days and light curves with periods greater than 0.4 days. The former are sinusoidal, while the latter are triangular and asymmetric. In the {\it W}1- and {\it W}2-bands, the light curve shapes are  homogeneous, with no appreciable difference between the two subclasses. \label{per_binned}}
\end{center}
\end{figure}

Figure \ref{per_binned} demonstrates the evolution of light curve shape as a function of period in the four wavebands. The space of possible periods for RR Lyrae stars, 0.2 to 1 day, was divided into six period bins: 0.2 to 0.3 day, 0.3 to 0.35 day, 0.35 to 0.4 day, 0.4 to 0.6 day, 0.6 to 0.7 day, and 0.7 to 0.9 day. Sources that fall into the first three bins are mostly subclass RRc stars and sources that fall into the last three bins are mostly subclass RRab, although this period boundary is not strictly defined. All of the light curves in the database are normalised to attain their minimum magnitude at 0 and maximum magnitude at 1. The normalised light curves are then grouped according to period bin. Each coloured curve represents the average of all light curves in the corresponding period bin. In the {\it V}- and {\it I}-bands, there are two distinct groups: the light curves in the first three period bins and in the last three period bins group together. However, in the {\it W}1- and {\it W}2-bands, the two groups vanish, and the average light curves for all of the period bins become undistinguishable.

\begin{figure}[h!]
\begin{center}
\includegraphics[width=1 \textwidth]{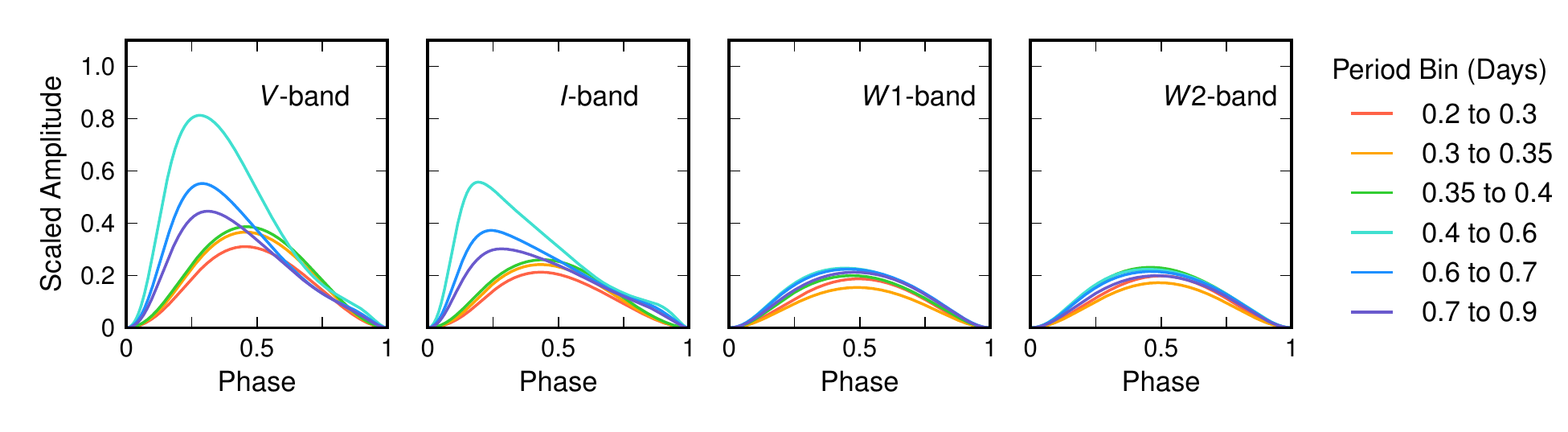}
\caption{Average light curves in Figure \ref{per_binned} are shown scaled by the mean amplitude of all light curves in the period bin. In {\it I}- and {\it V}-bands, RRab and RRc subclasses are distinguished by their amplitudes, while in the {\it W}1- and {\it W}2-bands, there is no significant variation in amplitude within the subclasses. \label{per_binned_amp}}
\end{center}
\end{figure}

Figure \ref{per_binned_amp} adds a visual representation of the average amplitude of the period bins. Each coloured light curve, as presented in Figure \ref{per_binned} with a normalised flux amplitude of 1,  is now scaled by the average amplitude of all the light curves in the period bin. While Figure \ref{per_binned} only provides information about light curve shape, Figure \ref{per_binned_amp} incorporates information about the light curve amplitude. Both of these features evolve from optical to mid-infrared wavelengths. 

\begin{figure}[h!]
\begin{center}
\includegraphics{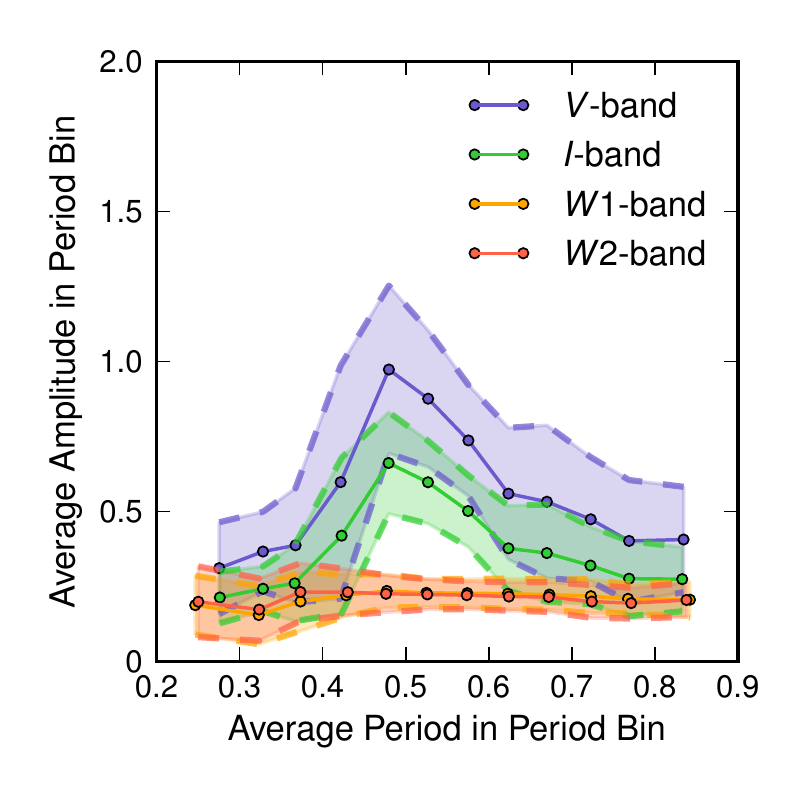} 
\caption{Binned period-amplitude plots for OGLE-III ({\it V}-band and {\it I}-band) and WISE ({\it W}1-band and {\it W}2-band) light curves. The period bins used are 0.2 to 0.3, 0.3 to 0.35, 0.35 to 0.4, 0.4 to 0.45, 0.45 to 0.5, 0.5 to 0.55, 0.55 to 0.6, 0.6 to 0.65, 0.65 to 0.7, 0.7 to 0.75, 0.75, 0.8, and 0.8 to 0.9 day. The average amplitude of all light curves in the period bin is drawn as a function of the average period of all light curves in each period bin. One standard deviation is drawn as the dashed coloured lines. The OGLE-III light curves have a clear maxima in the 0.45 to 0.5 bin, while the WISE plots remain relatively flat across the period bins. \label{bin_scatter}} 
\end{center}
\end{figure}

The information of Figure \ref{per_binned_amp} is further summarised in Figure \ref{bin_scatter}. Here, the number of period bins is increased to 12 in order to provide a meaningful plot, and the average amplitude of light curves in each period bin is plotted as a function of the average period for each bin. At shorter wavelengths, light curves with periods of around 0.45 day exhibit the largest amplitude, while in mid-IR wavelengths, all light curves express roughly the same amplitude regardless of period. 

In optical bands, the RRab and RRc subclasses of RR Lyrae stars are defined by the amplitude of the light curve and the light curve shape. However, in the mid-IR, these two classes appear less distinct, and accurate subclass classification based solely on mid-IR photometry becomes difficult. The reason for this distinction in optical light curves is due to the pulsation mechanism of RR Lyrae stars. During pulsation three different properties of the star are changing: the bolometric magnitude, the radial velocity, and the temperature \citep{1994ApJS...93..233B}. Observing a star at different wavelengths probes different stellar properties. The optical light curve from an RR Lyrae star is strongly affected by the temperature variation, whereas the mid-IR light curve is dominated by the variation in radius. This results in a triangular feature observed in longer-period optical light curves, corresponding with the RRab subclass. Since the mid-IR light curves are more strongly affected by the harmonic radial oscillation, the triangular feature is diminished at these wavelengths and the shape of these curves is more symmetric.

\section{Mid-IR Templates}\label{template_analysis}

\subsection{Procedure for Constructing Templates}
The next goal was to create a set of templates that represents typical light curve shapes in the WISE RR Lyrae catalog. A visual investigation of entire catalog reveals that there are several key features that distinguish light curves. These features include the slope of the light curve before it attained its maximum (in the region approximately 0.1 to 0.2 in phase space), the phase location of the maximum, the full width at half maximum, and the amount of symmetry in the light curve. The basic procedure was to group light curves based on similarity of shape, and then take the average of each group of light curves as a template for that group.

A challenge to the application of this methodology is the possibility that light curve shapes naturally form a continuous spectrum. For any feature, there is a maximum and minimum value, but any value between these is allowed. For example, suppose that you want to make light curve templates using only one feature: the phase location of the maximum brightness. Some light curves are left leaning and some are right leaning, and it is not difficult to imagine that there exists a light curve with any location of maximum brightness between the two extremes. Template clustering on this single feature amounts to picking N values for the feature between the maximum and the minimum, and grouping each light curve into a bin. The situation is complicated by the necessity of constructing these clusters based on all the features simultaneously, not just a single feature. 

The next challenge regards the optimal number of templates to produce. If a light curve lies halfway between two bins, it will be a good candidate for either bin. Employing many bins eliminates some ambiguity, but using fewer bins will make future analysis more meaningful. It was found that 10 to 15 bins is optimal.  

To ensure that the set of templates is representative of true RR Lyrae light curve shapes, a sample of high quality light curves from the full WISE RR Lyrae catalog was selected for the template construction process. The sample was created by selecting light curves with a minimum of 16 observations, a maximum robust model error of 0.04 magnitudes (see Section \ref{robust}), and a maximum gap in phase coverage of 15\%, which is 0.94 radians. A total of 1633 {\it W}1 light curves and 840 {\it W}2 light curves was used. 
The main feature of interest for this process was light curve shape, since this is an intrinsic property of the source while the mean magnitude is dependent on the distance at which observations are made. For this reason, light curves were normalised to achieve maximum magnitude at 0 and minimum magnitude at 1. 

Agglomerative hierarchical clustering \citep{linkage.cluster} was used to group the light curves in the sample. First, a distance between each pair of light curves was computed using the Euclidean metric. Hierarchical clustering was used to create a dendrogram from the condensed distance matrix. In other words, each light curve was initially grouped into its own cluster and the algorithm repeatedly combined the two closest clusters, eventually resulting in a single cluster containing the entire set of elements. The linkage criteria used in this step was average linkage clustering, which is to say that distance between any two clusters was determined by the average of all distances between pairs of objects in the clusters. The clustering algorithm was set to stop merging clusters once 23 nodes had been formed. Next, any cluster in the {\it W}1- band containing less than 15 sources and any cluster in the {\it W}2- band containing less than 10 sources was discarded. This procedure was used on both WISE bands and resulted in 10 clusters for each band. The average of each cluster was taken to be one light curve template.

\begin{figure}[h!]
\centering
\parbox{7.8cm}{
\includegraphics[width=1\linewidth]{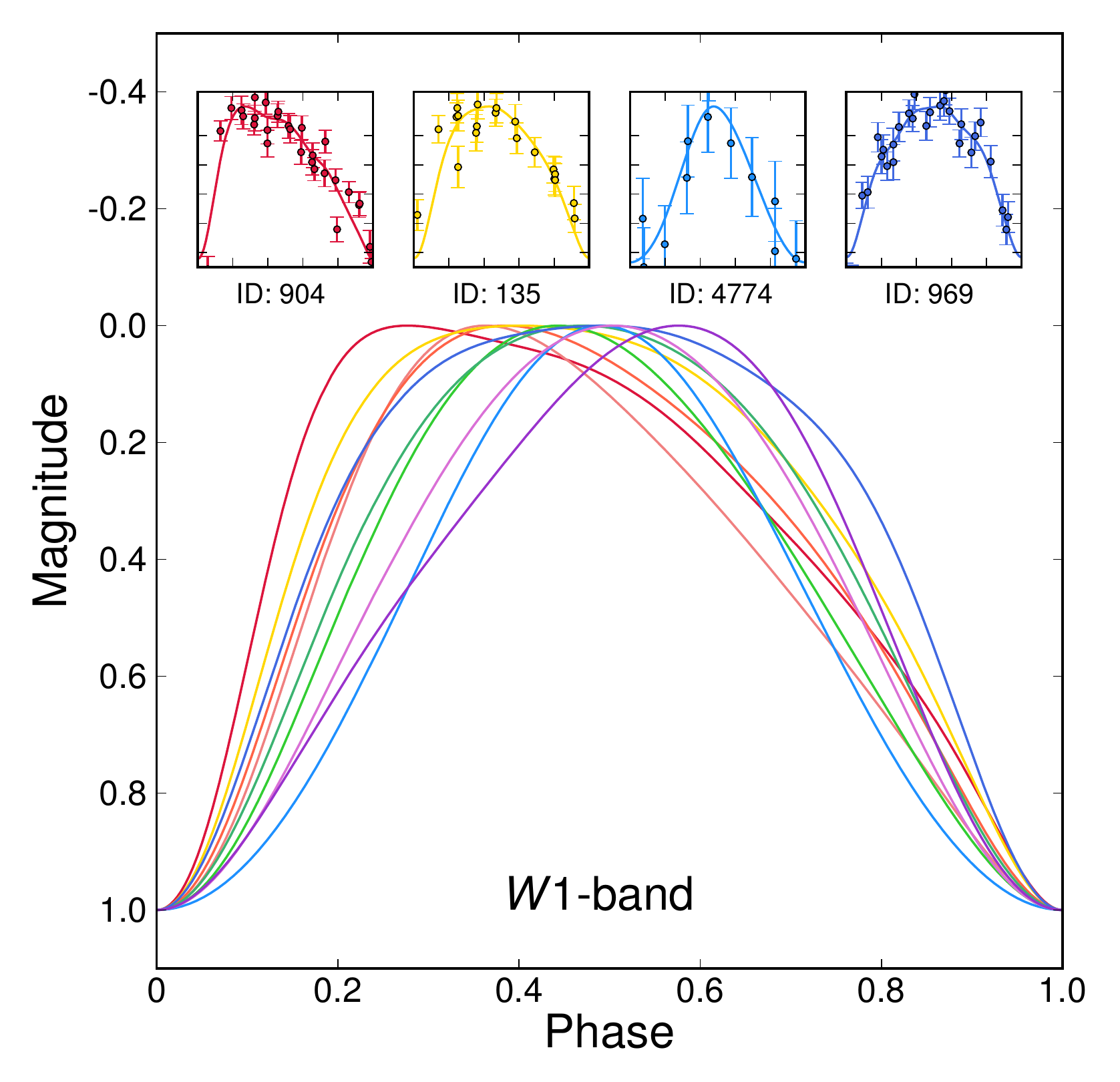}
\label{}}
\qquad
\begin{minipage}{7.8cm}
\includegraphics[width=1\linewidth]{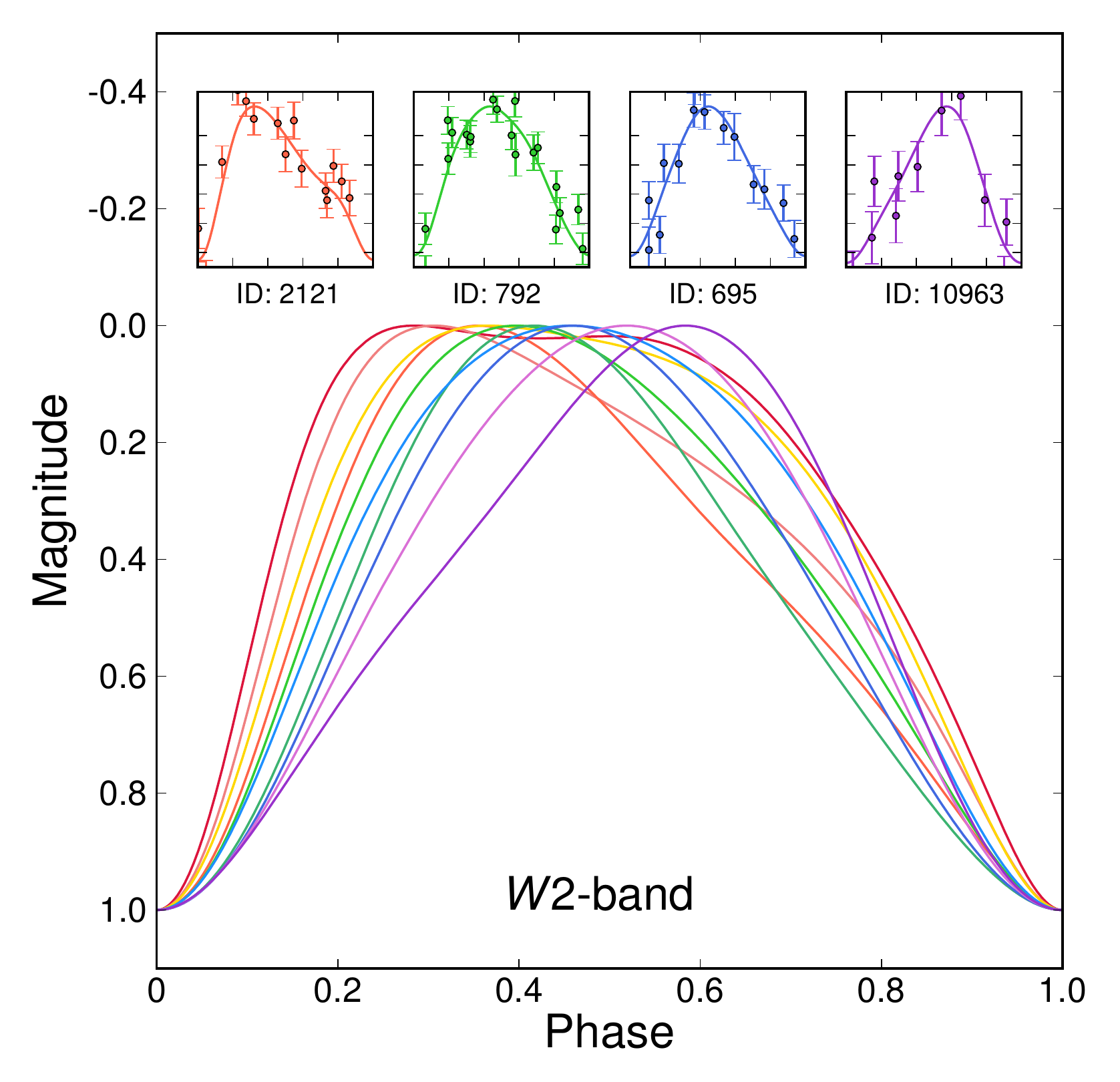}
\label{}
\end{minipage}
\caption{{\it W}1- and {\it W}2-band light curve templates resulting from hierarchical clustering methods on light curve shape. The template colour is based on the relative location of maxima, with red indicating the template with the leftmost maximum and purple indicating template with the rightmost maximum. The embedded plots show real light curves from the corresponding template. For these small plots, the light curve spans one full phase cycle and the amplitude is scaled to be 0 at maximum magnitude and 1 at minimum magnitude.}
\label{temps}
\end{figure}

Figure \ref{temps} shows the 10 templates created for each of the {\it W}1- and {\it W}2-bands. For each band, four sample light curves are displayed, colour--coded by their associated template. Each set of templates appears consistent with a continuous distribution of light curve shapes. The main feature that varies throughout the set is the phase location of the maximum brightness, but this is not the only feature useful for distinguishing light curve shapes. In the {\it W}1-band, templates 8 and 9 have a very close location of maxima (at 0.494 and 0.504 in phase space, respectively) but they have a different value of the full width at half maximum (0.698 and 0.553, respectively). The complexity of the light curve shape provides justification for the method of hierarchical clustering on the whole template shape, as opposed to a single feature. The templates are provided in electronic format along with the other supplemental material. 

\subsection{Using Templates for Mean Magnitude Estimation}
A practical application of light curve templates is the recovery of light curve information for poorly-sampled sources. Suppose a source with a known period is sampled a handful of times in the {\it W}1- or {\it W}2-band. The objective is to estimate the mean magnitude, which is needed to apply period--luminosity relations for the distance measurement. The most straightforward way to estimate the mean magnitude is to average all of the magnitude measurements in the light curve. However, this result may be inaccurate due to sampling biases and asymmetries in the light curve. A more precise method is to perform a Fourier analysis and fit a harmonic model to the light curve, then take the mean magnitude of the model. This method may result in over--fitting, or fitting extra bumps and dips that are not representative of the true light curve. Over--fitting is especially likely if the light curve is poorly sampled, has gaps in phase coverage, or has a low signal--to--noise ratio. An alternative method is template fitting: for each normalised template in a class-- and bandpass--specific catalog, scale the period, amplitude, and minimum magnitude in order to minimise the residual difference between the light curve and the template. The template with the lowest residual is chosen as the best match and the mean magnitude of that template fit is taken to be the mean magnitude of the source.

To demonstrate the effectiveness of the three mean magnitude estimation methods (straight mean, harmonic model mean, and template mean), each method was used on a sample of artificial light curves, varying the number of epochs of observation (N) and the photometric sensitivity ($\sigma$). First, a random model light curve was selected from the full WISE RR Lyrae catalog and resampled randomly at N positions. Noise was added to the light curve by taking a Gaussian sample of each magnitude datapoint with a standard deviation of $\sigma$, similar to the bootstrap perturbations used in Section \ref{robust}. An artificial light curve was created for the following parameter values: N = 5, 6, 7, 8, 9, 10, 11, 12, 13, 14, 15, 20, 25, 30, 40, 50, 60 and $\sigma$ = 0.025, 0.05, 0.075, 0.1, 0.15 magnitudes. The mean magnitude of each artificial light curve was estimated using the three methods described above. For each value of N and $\sigma$, 500 artificial light curves were generated.

\begin{figure}[h!]
\begin{center}
\includegraphics{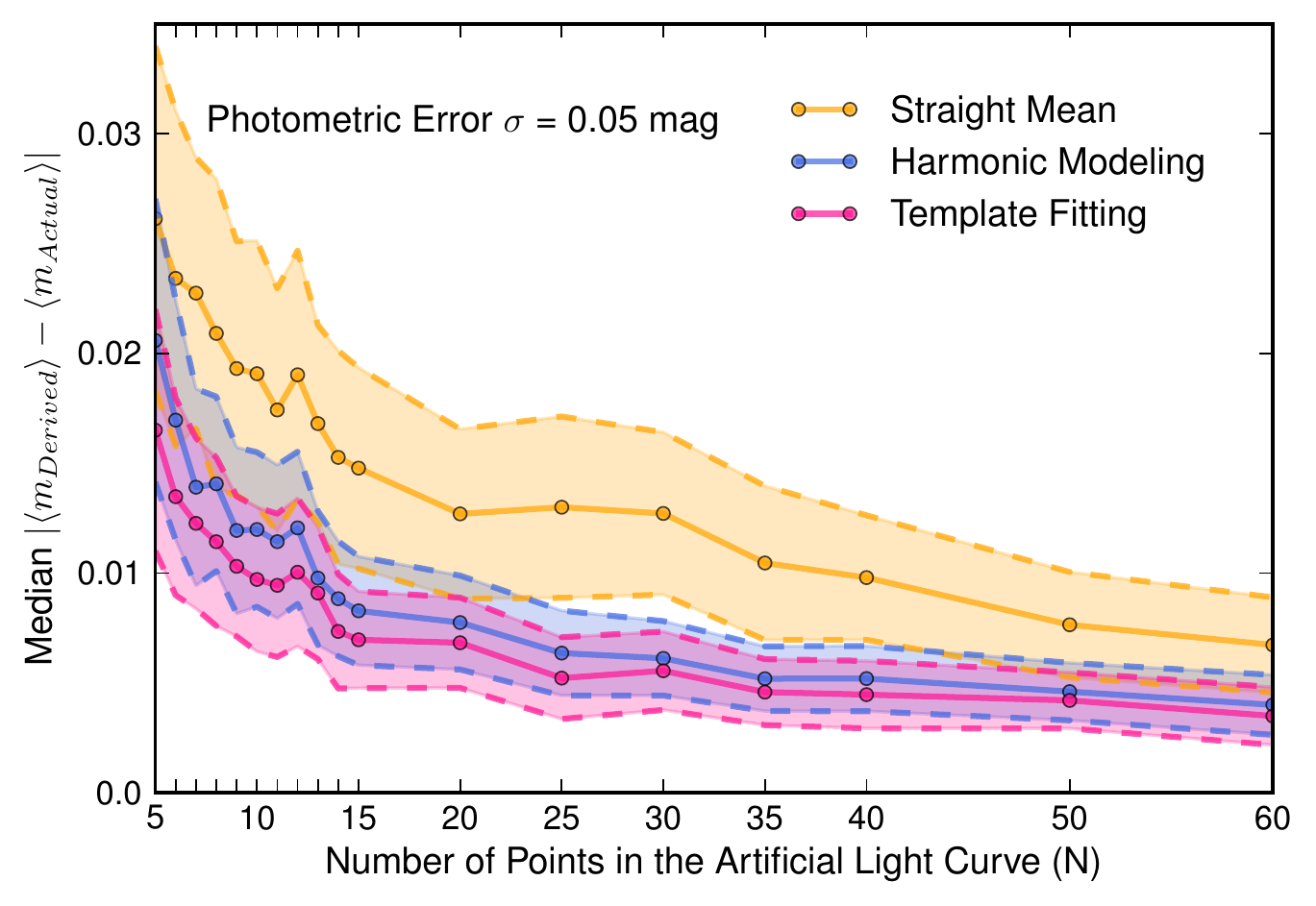}
\caption{ A comparison between the accuracy of three methods for deriving the mean magnitude of a light curve. For each number of light curve data points N, 500 random light curves from the WISE RR Lyrae database were selected and sampled randomly at N points. An artificial light curve was created by adding noise to the magnitude value, taking a Gaussian magnitude sample with the magnitude value for the mean and $\sigma$ for the standard deviation. The mean magnitude of the artificial light curve was recovered with each of the three methods, and a mean magnitude residual was calculated for each method. Each filled circle represent the median values of each set of mean magnitude residuals, with dashed lines drawn one median absolute deviation.} \end{center}
\label{comp_plot}
\end{figure}

For each of the three methods, for each value of N and $\sigma$, a mean magnitude residual was calculated by taking the absolute value of the difference between the recovered mean and the actual mean of the model. The method that most accurately predicts the mean magnitude of the light curve will have the lowest overall mean magnitude residual. Figure \ref{comp_plot} shows the median and median absolute deviation for each mean magnitude estimation method, keeping $\sigma$ fixed at 0.05 magnitude.

As seen in Figure \ref{comp_plot}, template fitting generally provides more accurate mean magnitude values than harmonic modelling. When the number of data points is larger than 30, the difference between the two methods becomes negligible. The straight mean has the largest scatter and the lowest accuracy and should not be used for mean magnitude estimates, even with a high number of data points. These findings are consistent with \citet{2007A&A...462.1007K}, despite significant differences in methodology used for template creation. 

The templates can be used as a direct application for mean magnitude estimates for light curves with less than 30 points observed with JWST. If measuring the distance to an RR Lyrae is the goal of observing the source, then the only information needed is the mean magnitude of the source. Templates make it possible to recover this measurement with the same accuracy as harmonic modelling, but requiring fewer observations, essentially reducing the necessary telescope time. When the given templates are not directly applicable because of the bandpass used is not the {\it W}1- or {\it W}2-band, the methodology is transferrable. While the process of generating a complete set of light curves requires an initial sample of well-observed light curves, it will cut down observation time for the rest of the light curves. It may be economical to invest observation time to generate a templates set and then use the set to obtain accurate mean magnitude estimates for all subsequent sources with fewer epochs of observation. 

\section{Conclusions}\label{conclusion}
We have presented a catalog of 3740 RR Lyrae stars in WISE with derived periods and distances. This catalog can be used for creating dust maps of the galaxy up to distances of 14 kiloparsecs. Most certainly the catalog does not contain every RR Lyra star observed by WISE. However, the sample is large enough that it summarises the typical shapes of mid-infrared RR Lyrae light curves. The catalog may be used as a training set in a machine learning initiative to identify all RR Lyrae stars in the full WISE database.

Harmonic light curve models were created for the sources in the catalog. To study light curve morphology, these models were compared with similarly--derived harmonic models from the OGLE-III catalog of variable stars. There are two main distinctions between the models in OGLE-III bands and those in WISE bands. First, light curve amplitudes are higher for visible wavelengths. For the {\it V}-, {\it I}-, {\it W}1- and {\it W}2-bands, the mean RRab amplitudes are 1.147, 0.595, 0.228, and 0.222 magnitudes and the mean RRc amplitudes are 0.546, 0.279, 0.162, and 0.186 magnitudes. The second difference is that models in visible wavelengths display significant variation in light curve shape depending on the source period, while models in the mid-infrared have rather uniform light curve shapes. This trend in light curve morphology is depicted in Figure \ref{per_binned}. Evolution in shape as a function of bandpass arises because different bandpasses probe different stellar layers. Observations at long wavelengths yield information about the relationship between the period and the radius of the star, which undergoes harmonic oscillation. Observations at shorter wavelengths are sensitive to temperature variation, which exhibits sawtooth--like patterns.

We have created 10 templates for typical light curve shapes in the {\it W}1- and {\it W}2-bands, employing statistical clustering methods. We have investigated the application of these templates as a method for the estimation of mean magnitude. Template fitting generally improved on mean magnitude estimates from harmonic modelling methods, but the level of improvement was negligible when used on sources with over 30 data points. With a simulated photometric sensitivity of 0.05 magnitudes, harmonic modelling on a light curve with 25 observations will typically estimate the mean magnitude to within 0.006 magnitudes of the actual value. The same accuracy can be achieved using template fitting on a light curve with 18 observations. This improvement may become significant during large RR Lyrae observation campaigns. 

\vspace{18 mm}

{\bf Acknowledgements}: We thank Nathaniel R. Butler for providing the implementation of the Lomb-Scargle algorithm that was used in this paper. This publication makes use of data products from the Wide-field Infrared Survey Explorer, which is a joint project of the University of California, Los Angeles, and the Jet Propulsion Laboratory/California Institute of Technology, funded by the National Aeronautics and Space Administration. This research has made use of the SIMBAD database, operated at CDS, Strasbourg, France.

\clearpage

\bibliography{Klein_refs}

\end{document}